# A Deep Learning Pipeline for Large Earthquake Analysis using High-Rate Global Navigation Satellite System Data

Claudia Quinteros-Cartaya[1], Javier Quintero-Arenas[1,2], Andrea Padilla-Lafarga[3], Carlos Moraila[3], Johannes Faber[1,4], Wei Li[1], Jonas Köhler[1,5], Nishtha Srivastava[1,5*]

[1]Frankfurt Institute for Advanced Studies, Ruth-Moufang-Straße 1, Frankfurt am Main, 60438, Hessen, Germany.
[2]Institute of Computer Sciences, Goethe University, Robert-Mayer-Str. 11-15, Frankfurt am Main, 60325, Hessen, Germany.
[3]Faculty of the Earth and Space Sciences, Autonomous University of Sinaloa, Universitarios Ote. S/N, Cd Universitaria, Culiacan Rosales, 80040, State, Mexico.
[4]Institute for Theoretical Physics, Goethe University, Max-von-Laue-Str. 1, Frankfurt am Main, 60438, Hessen, Germany.
[5]Institute of Geosciences, Goethe University, Altenhöferallee 1, Frankfurt am Main, 60438, Hessen, Germany.

*Corresponding author(s). E-mail(s): srivastava@fias.uni-frankfurt.de;
Contributing authors: quinteros@fias.uni-frankfurt.de;
jaqarenas@fias.uni-frankfurt.de; andreapadilla@uas.edu.mx;
cmoraila@uas.edu.mx; faber@fias.uni-frankfurt.de;
wli@fias.uni-frankfurt.de; jkoehler@fias.uni-frankfurt.de;

**Abstract**

Deep learning techniques for processing large and complex datasets have unlocked new opportunities for fast and reliable earthquake analysis using Global Navigation Satellite System (GNSS) data. This work presents a deep learning model, MagEs, to estimate earthquake magnitudes using data from high-rate GNSS stations. Furthermore, MagEs is integrated with the DetEQ model for earthquake detection within the SAIPy package, creating a comprehensive pipeline for earthquake detection and magnitude estimation using HR-GNSS data. The MagEs model provides magnitude estimates within seconds of detection when using stations within 3 degrees of the epicenter, which are the most relevant for real-time applications. However, since it has been trained on data from stations up to 7.5 degrees away, it can also analyze data from larger distances. The model can process data from a single station at a time or combine data from up to three stations. The model was trained using synthetic data reflecting rupture scenarios in the Chile subduction zone, and the results confirm strong performance for Chilean earthquakes. Although tests from other tectonic regions also yielded good results, incorporating regional data through transfer learning could further improve its performance in diverse seismic settings. The model has not yet been deployed in an operational real-time monitoring system, but simulation tests that update data in a second-by-second manner demonstrate its potential for future real-time adaptation.

# 1 Introduction

Although deep learning (DL) algorithms have shown considerable potential in earthquake analysis, their use for seismic monitoring with high-rate Global Navigation Satellite System (HR-GNSS) data remains relatively underexplored. During the past few decades, HR-GNSS data have gained recognition as valuable complementary information for earthquake analysis (e.g. [26], [11], [19], [23], [4], and [7]). Despite the inherent advantages of GNSS data, such as its ability to measure large ground displacements without saturation, the application of DL algorithms to this field remains a significant challenge. The integration of DL into GNSS seismology is still in its early developmental stages, with relatively few studies in recent years (e.g. [15], [14], [22], [2]).

Among the primary challenges are the limited availability of real HR-GNSS data for large earthquakes to train robust models, the inherent noise in GNSS signals, and the need for DL models to generalize effectively across diverse tectonic regions. Furthermore, achieving efficient event detection followed by rapid and reliable seismic source analysis remains a significant challenge for real-time seismic monitoring.

In this paper, we propose a novel DL model, MagEs, tailored to estimate the magnitude of large seismic events through ground displacement data. The proposed model is based on a model architecture of a Residual Network (ResNet-like) [6]. We detail its evaluation using a dataset comprising HR-GNSS data from both synthetic and real seismic events. We evaluated the performance and robustness of our model through experiments with datasets from various seismogenic regions. We explore diverse dataset configurations, including epicentral distances, signal duration, and earthquake magnitudes, to evaluate the adaptability and generalization of our model in different seismic scenarios.

Furthermore, our proposed model, MagEs, is integrated with DetEQ, an earthquake detection model [21], to create an additional branch within the SAIPy package [12]. This integration enables the identification of events from displacements recorded by multiple GNSS stations and facilitates an estimate of the magnitude within seconds after detection at each station. This comprehensive pipeline, which draws on HR-GNSS data, constitutes a complementary branch for seismic analysis within the SAIPy package.

## 2 MagEs: Magnitude Estimation

### 2.1 Architecture

Our proposed deep learning model, which we call MagEs, is a residual neural network (ResNet-like) for regression prediction, constituted by convolutional layers in residual blocks, pooling layers, and a final fully connected layer (Figure 1).

In broad terms, MagEs is designed to process multivariate time series data with an input shape of (3 x 390 x 3), representing data from three distinct receivers (GNSS stations), each with a window of 390 time-steps and three channels per receiver (latitude, longitude, and height coordinates). For the specific problem addressed in this work, the configuration of the input data is described in detail in the following section (see Figure 2).

The model includes six residual blocks, whose connections help propagate the gradient through the network during training, mitigating the vanishing gradient problem [6].

Each residual block consists of two branches, one with three sequential convolutional layers and another with one convolutional layer. The sequential convolutional layers are regularized using the L2 regularizer, with a scaling factor $\lambda$ = 5e-4 (determined empirically), which multiplies the regularization term to penalize large weights and prevent overfitting [5]. In contrast, the convolutional layer in the skip connections is not regularized, allowing essential information to pass through the model without excessive constraints.

The number of filters, determined through experimentation, increases progressively across the blocks: 32, 64, 128, 256, 512, and 1024. All convolutional layers use a kernel size of (1, 3), with a Glorot normal initializer for weight initialization, and LeakyReLU as the activation function to introduce non-linearity [3], [5].

The output of each residual block is computed by adding the element-wise sum of the outputs from the sequential convolutional layers and the output from the convolutional layer in skip connection. This addition allows the model to learn the residual, enabling it to generalize more effectively to new data rather than learning a full transformation from the input.

After the first five residual blocks, a max-pooling layer with a pool size of (1, 2) is applied, halving the number of time-steps and enabling the model to focus on the most significant temporal patterns. After the final residual block, a global-max-pooling layer (1,1) is applied instead of a typical pooling layer, compressing the time series data into a single vector representation of size (1, 1024), corresponding to the number of filters in the last convolutional layer.

Following the residual blocks, a dense layer with 128 neurons is applied, using the LeakyReLU activation function. In this layer the weights are regularized to prevent overfitting, constrained by a maximum norm value of 3 (determined empirically) and with a Glorot normal initialization [3], [5]. The final output layer produces a single floating-point value (rounded to one decimal place), representing the magnitude of an earthquake in our particular application.

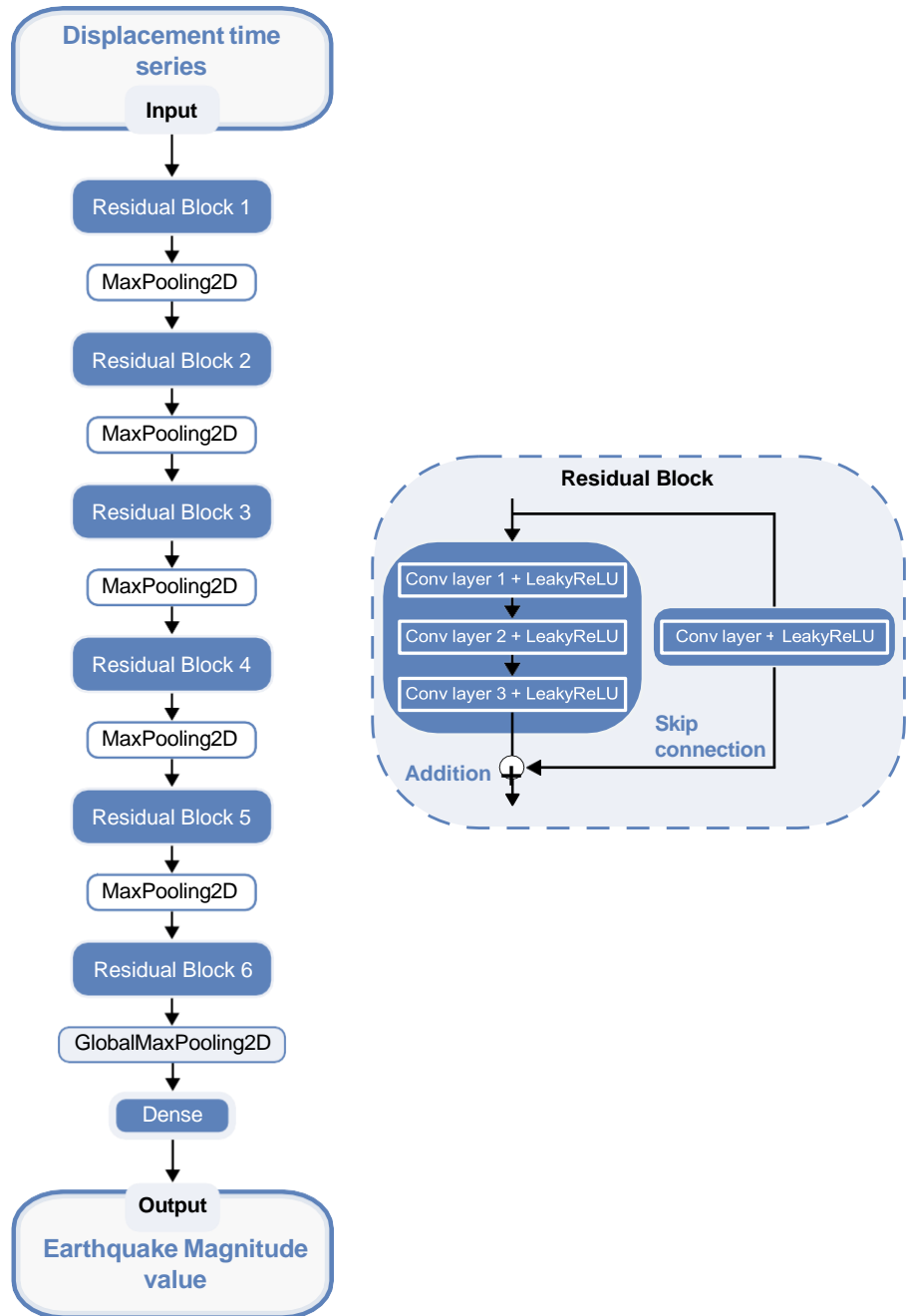

**Fig. 1** Sketch of the architecture of MagEs model, which consists of a residual neural network (ResNet-like). In the right, the structure of each residual block in the network is illustrated in detail.

## 2.2 Input Settings

The model works on time series that represent the ground displacement observations by GNSS stations, in physical meters units (not normalized), recorded at a 1 Hz sampling rate. The start time of the time series is the second at which the first local ground displacement caused by the earthquake is detected at the station. The maximum duration of the time series allowed by the model is 390 seconds and is designed to work with data from up to three stations at the same time. For each station, three channels are considered: North, East, and Vertical, corresponding to observations in the latitude, longitude, and height coordinates. Consequently, the input data are organized into a tensor with dimensions (3 x 390 x 3), which represents the maximum number of stations, the maximum length of the time series, and the three channels (Figure 2).

The flexibility concerning the number of stations and length of the time series is handled by zero padding to maintain a consistent input shape. That is, in cases where fewer than three stations are utilized, the lack of data from the remaining stations is padded with zeros. Likewise, when the time windows are <390 seconds, the end of the time series is padded with zeros as well.

The variety in the length of the time series (before padding) is because we determined the length of the time series considering that the duration of the local ground displacements caused by earthquakes depends on the epicentral distance. Hence, we defined a linear function

$$t_w(x) = t_0 + [m * x], \qquad (1)$$

in such a way, that by using a slope straight line $m = 50 s/° = 0.45 s/km$ (Figure 3), the minimum time window is $t_0 = 15$ seconds for distances of 0° and a maximum window of 390 seconds corresponds to the farthest epicentral distance included in the dataset, 7.5°( 834 km).

## 2.3 Data

The dataset used in this study comprises synthetic data and real data. Synthetic data are employed to address the limited availability of GNSS observations from large earthquakes, which are insufficient to effectively train the DL model. In contrast, real data are used to evaluate the performance of the model under real-world conditions. The following subsections provide a detailed description of each dataset.

### 2.3.1 Synthetic Data

Due to the limited real HR-GNSS data from large earthquakes to train the model, which typically requires thousands of data for effective training, we used a synthetic dataset, with magnitudes ranging from 6.9 ≤ Mw ≤ 9.6, generated from idealized earthquake scenarios. This dataset consists of displacement time series observed by various GNSS stations, resulting from theoretical ruptures in the Chilean subduction zone, generated by Lin et al. (2021) [13], [14].

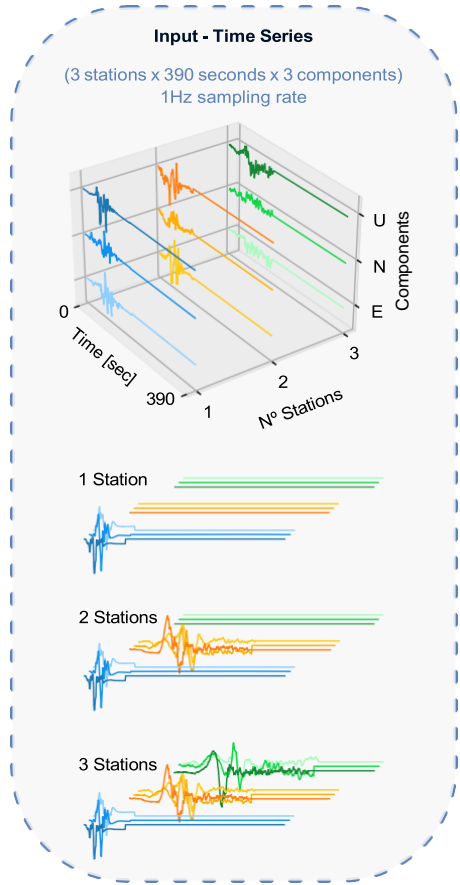

**Fig. 2** The input data consists of a tensor constituted by those time series from a maximum of three stations, with 390 time steps (seconds), and in three channels (North, East, and Vertical). In the case of having less than three stations, and depending on the time windowing of the data by epicentral distance (see Figure 3), the tensor dimensions of (3x390x3) are kept constant by zero padding.

Furthermore, for each signal, we introduced distinct white noise time series [18], [17], evaluating the signal-to-noise ratio to ensure that the noise amplitude did not mask the signal (see Figure 4 for signal and noise comparisons). This process allowed us to augment the dataset by including both noise-free signals and signals with different levels of noise.

We organized the data by magnitude, using 1, 2, and 3 stations, without repeating any data within the arrangements. The dataset was then split into training, validation, and test sets, allocating 90% of the data to training and validation, and 10% to testing. The training-validation set was further divided, with 80% used for training and 20% for validation.

Since the number of data for magnitudes between 7.2 and 9.4 was significantly higher compared to the other magnitudes (due to the proportionality in the original dataset), we imposed upper limits on the data for these magnitudes: 5000 for training, 3000 for validation, and 2000 for testing, per station configuration (Figure 5).

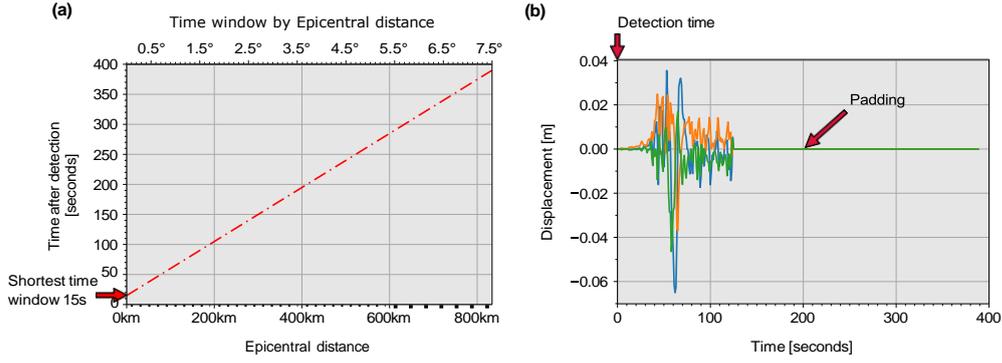

**Fig. 3** The length of each signal used for training was determined by the epicentral distance. (a) The relationship between signal length and epicentral distance is defined by Equation 1. The shortest signal is 15 seconds for a station located at 0 km from the epicenter, while the longest signal spans 390 seconds for a distance of 834 km. (b) An example of the three-channels signals from one station with an epicentral distance of 243.8 km is shown. The time series is padded to fill 390 seconds to be consistent with the input shape for the model.

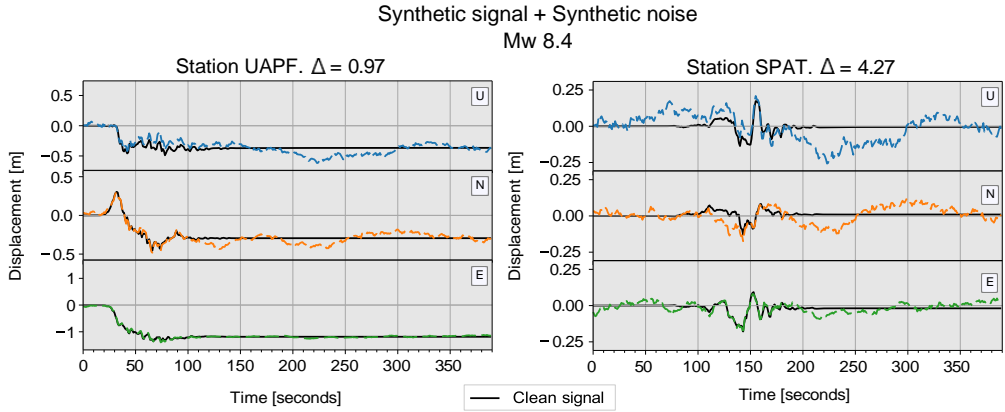

**Fig. 4** Comparison between the original synthetic signal (black solid line) and the signal with added synthetic noise (dashed lines) for an example earthquake of Mw 8.4. The example highlights how noise is more pronounced in the signal from the farthest station.

### 2.3.2 Real Data

We evaluated the performance of MagEs using two datasets with real GNSS observations. The first dataset comes from a previously published source compiled by Melgar and Ruhl (2018) [23], [16], and includes earthquakes from different regions with magnitudes (Mw) greater than 6.9 (Figure 6). This dataset was used to test the reliability and accuracy of the magnitude estimation under controlled conditions where the epicentral distance is known, and thus the time window length is defined by distance.

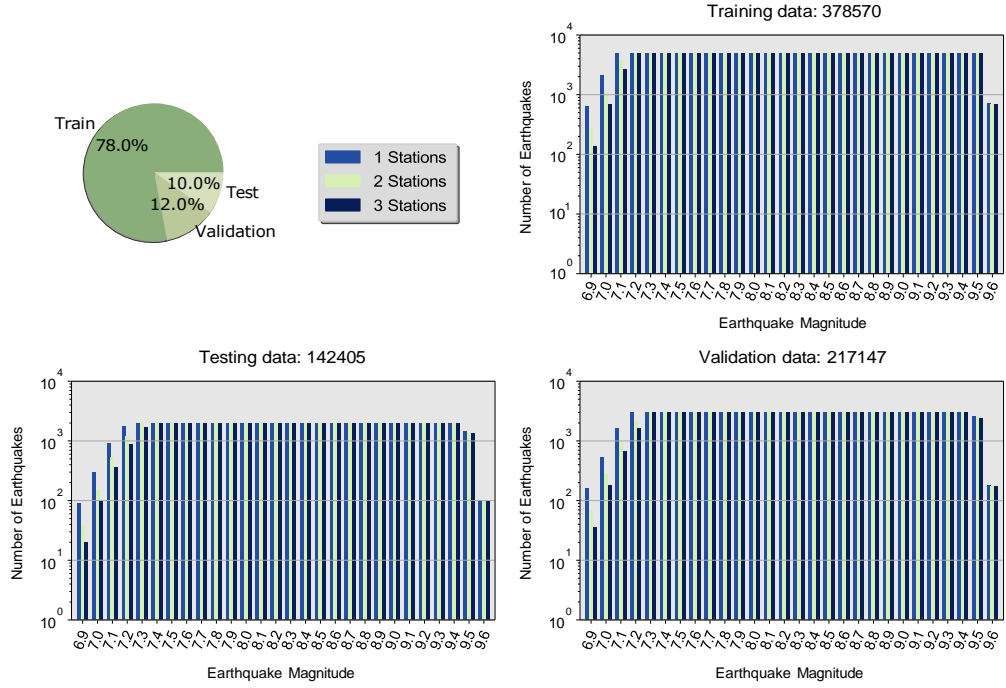

**Fig. 5** Data distribution for training, validation, and testing sets after data augmentation by adding noise. The dataset includes signals from one to three stations for each magnitude between 6.9 and 9.6. Data quantity, per number of stations in each magnitude, is limited to a maximum of 5000 samples for training, 3000 for validation, and 2000 for testing.

A second dataset contains continuous 24-hour GNSS observations, specifically used to test the full pipeline, including both the detection and magnitude estimation processes. This dataset was processed from RINEX files at a sampling frequency of 1 Hz [24], [20], [1]. The GNSS data processing techniques to obtain precise time series follows the same methodology applied by Quintero-Arenas et al. (2024) [21].

## 2.4 Training

The training dataset was labeled by the magnitude of the earthquakes, rounded to one decimal place, for a supervised learning process. To establish a practical baseline for zero input data while keeping the focus on labels with high magnitude values (range 6.9 to 9.6), we assign a minimum label of magnitude 6 for instances with zero input. Using this minimum value, rather than a label significantly below the training range (such as magnitude 0), helps prevent inconsistent or unstable model predictions. To ensure balanced training, we included zero input cases in the same proportion as other data points, with 5000 samples per case (per set of number of stations).

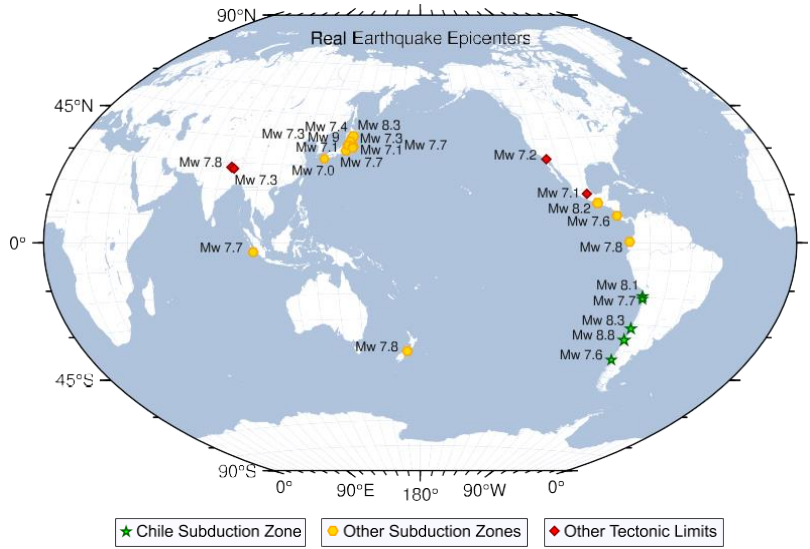

**Fig. 6** Earthquake epicenters of the real data used to test of MagEs model [16]

The training was optimized with the Adaptive Moment estimation (ADAM) algorithm [10], with a learning rate scheduler by the Cosine Restart Decay function, starting with a value of 1e-4 [5]. We fixed a batch size of 64, a maximum of 500 epochs, and an early stopping with patience of 20 epochs, once reaching the minimum validation loss.

The best training was obtained in the 55th epoch (Figure 7), where the loss value measured by the Mean Squared Error (MSE) was 0.0043 in the training and 0.0103 in the validation. Also, the Mean Absolute Error (MAE) during both the training and validation decreased to a minimum of 0.0500 and 0.0721, respectively (Figure 7).

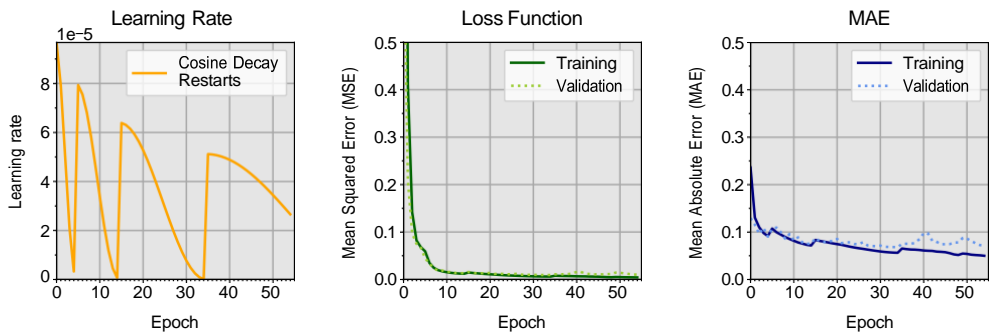

**Fig. 7** On the left, the Cosine Decay Restart function used to define the learning rate. In the center, the loss function measured by mean squared error for both training and validation. On the right, the mean absolute error recorded during training and validation.

## 2.5 Tests

Initial tests were performed using the synthetic dataset to evaluate the performance of the model with data defined under the same theoretical conditions as the training data.

Based on the number of stations used in our tests, we observed that the model performs effectively in all study cases when using data under ideal conditions. The model achieves root mean squared errors (RMS) of 0.12, 0.08, and 0.07, using one, two, and three stations, respectively (Figure 8). This level of precision indicates that the model consistently provides reliable magnitude estimates, even with variations in the number of stations.

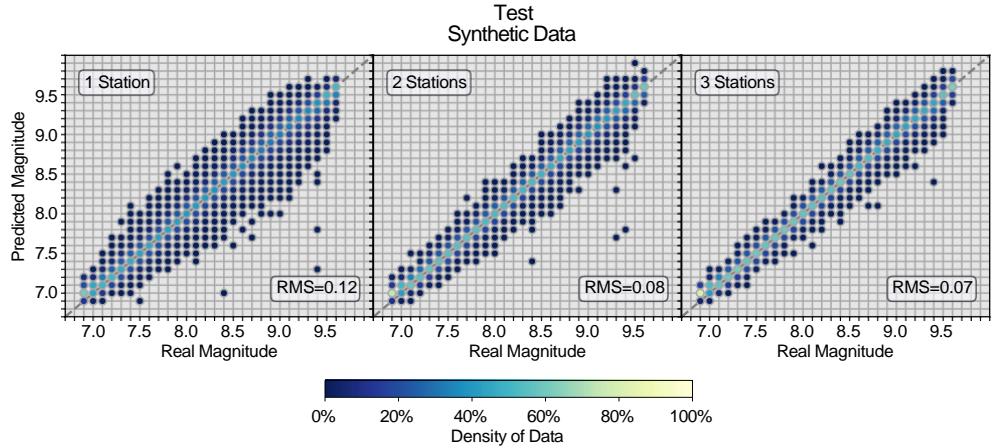

**Fig. 8** Results from testing with synthetic data using varying numbers of stations. The color scale illustrates that most results show a strong fit, with errors mostly concentrated at ≤ 0.1.

To evaluate the accuracy and reliability of the model under real conditions, we performed the evaluation using real data. As in the previous test, the data were assigned to one to three stations. In cases with multiple stations, the stations were randomly grouped to simulate various observation scenarios and assess how the combinations of the stations might influence magnitude estimates.

Since the model was trained using earthquake scenarios specifically from the Chilean subduction zone, our first evaluation with real data was conducted using data exclusively from Chile, to identify any potential limitations before applying the model to earthquakes in other regions with different tectonic conditions.

In Figure 9, we illustrate the results from the Maule earthquake data as an example, showing the distribution of magnitudes obtained through different combinations of stations as a function of the largest epicentral distance in each set of stations. This distribution highlights a non dependence of the accuracy and consistency of the magnitude values on distance.

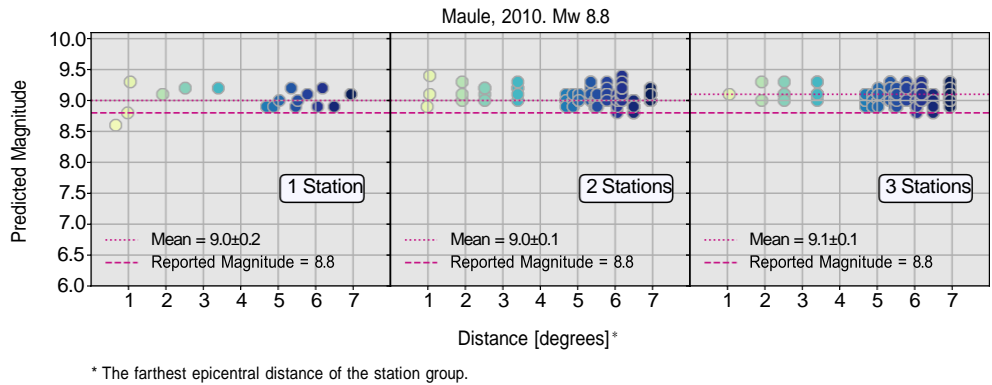

**Fig. 9** Example of test results using real data from the Maule earthquake, shown by the number of stations. For single station cases, the color scale represents the epicentral distance of each station. In the two and three station cases, station combinations are selected randomly, with colors and epicentral distances corresponding to the farthest station in each respective combination. Considering the mean of the magnitudes for each case, the results show an error of approximately 0.2.

In addition, Figure 10 shows the magnitude estimates for all Chilean earthquakes included in the database (epicenters in Figure 6). Here, the magnitude of each earthquake represents the mean of the estimates in each test case (from one to three stations). The results demonstrated a favorable performance of the model, as indicated by the RMS values of 0.15, 0.22, and 0.25 for each respective case. These low RMS values suggest that the model consistently estimates accurate magnitudes.

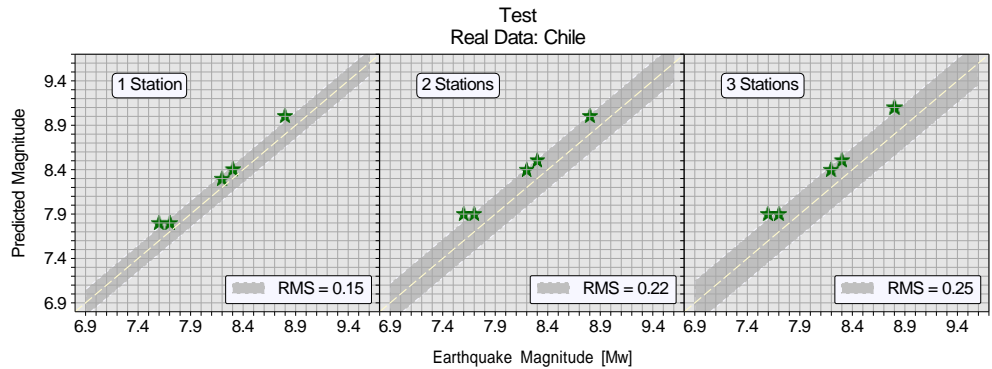

**Fig. 10** Results from testing with real data from various earthquakes in Chile. In all cases, a good fit is observed, with a slight overestimation of ≤ 0.3.

It is worth considering that noise interference or occasional signal loss can affect individual stations. Therefore, when MagEs uses data from groups of stations, these issues can also affect accuracy, as the estimation relies on the combined input of all included stations. This data quality problem might be reflected by an outlier in tests with a single station, while in tests with groups of two or three stations, it is not necessarily obvious.

In Figure 11 we illustrate the magnitude estimation second by second with different stations for five large earthquakes from Chile. In this example, we use stations within 7.5° epicentral distance. We can observe how the magnitude estimates tend to stabilize at values close to the expected magnitude at a time ranging from ∼30 seconds to two minutes, depending on the distance from the station and the quality of the data.

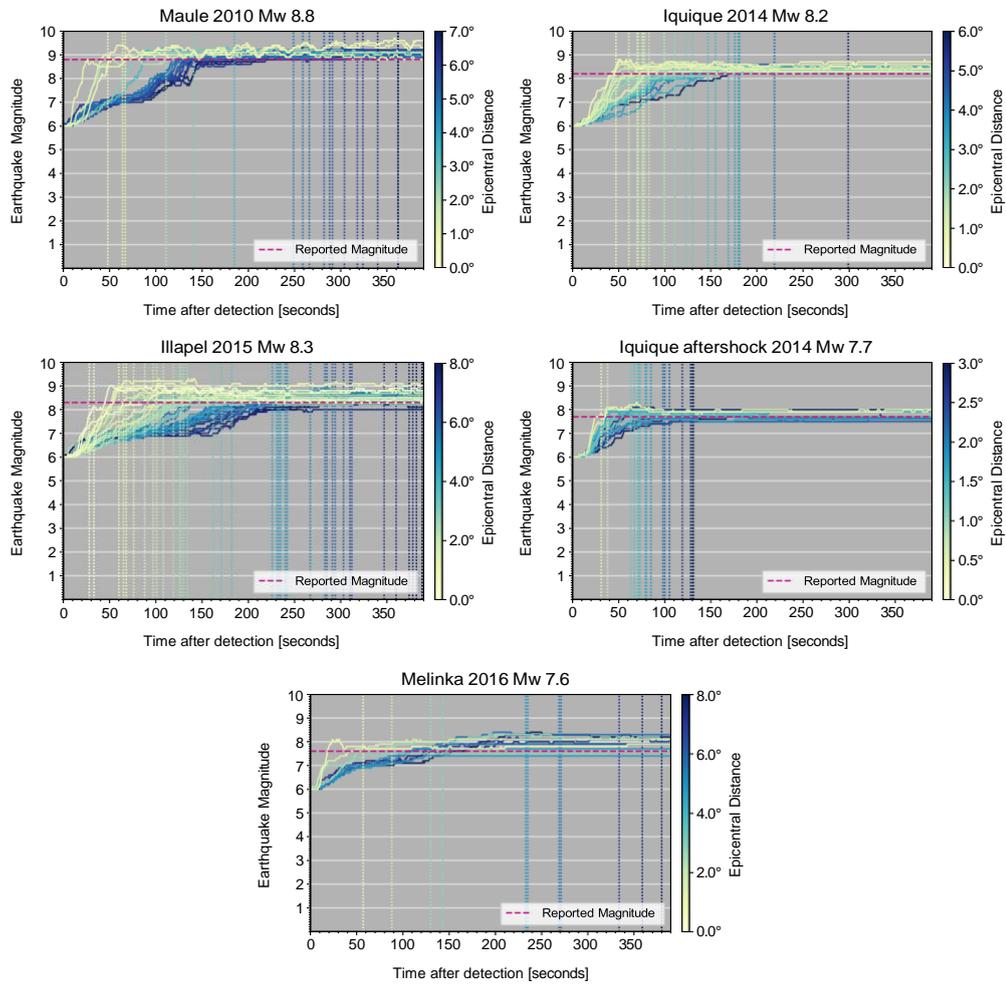

**Fig. 11** Test results with real data from earthquakes in Chile of varying magnitudes, showing the magnitude estimation as input data, and thus the output, is updated second by second. The color scale represents the epicentral distance of stations, limited to those within 7.5° for these tests. Dotted vertical lines mark the end of the training window for the respective epicentral distance, indicated by the color scale.

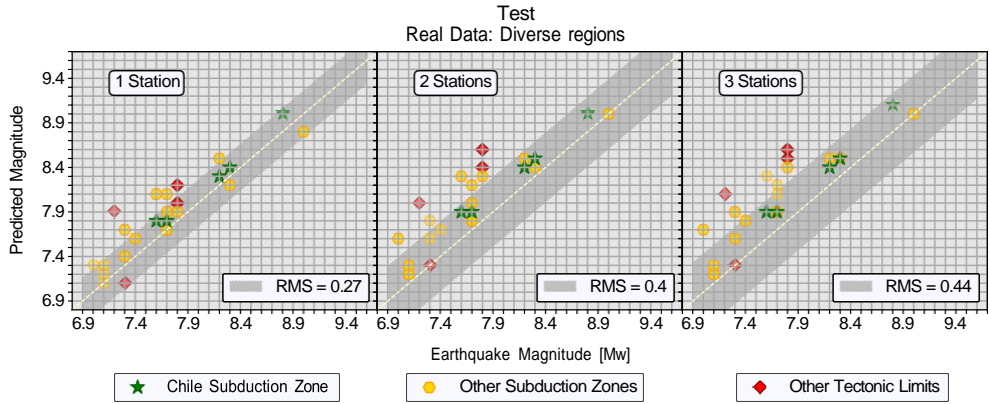

**Fig. 12** Results from testing with real data from diverse tectonic regions. Overall, most earthquakes exhibit a good fit, with the best results observed for earthquakes in Chile. While several earthquakes from other regions also show a very good fit, only a few cases, primarily from non-subduction zone events, have errors exceeding 0.5.

Posteriorly, we use real input data from diverse regions to gauge the capability of the model to generalize to distinct real seismogenic scenarios. These results again show that the lowest RMS is achieved when evaluating the mean magnitude of individual stations; the RMS was 0.27 for single-station estimates, 0.40 for two-station combinations, and 0.44 for three-station combinations (Figure 12). The highest errors in the experiment with combinations of 2 and 3 stations could be due to the influence of data quality. If a station with poor data quality is included in multiple combinations, the average of magnitude is affected. In general, the results suggest that the model can perform well across different geographic areas, in addition to verifying how well the training translates to real events in the same tectonic context.

# 3 Pipeline into SAIPy package

We incorporate a comprehensive pipeline for HR-GNSS seismic data analysis that integrates both DetEQ [21] and MagEs models into the SAIPy package [12]. This integration forms a workflow within SAIPy, allowing automated event identification and magnitude estimation (Figure 13).

The pipeline begins with DetEQ, an earthquake detection model that processes the three channels of data from multiple GNSS stations. DetEQ identifies the initial seismic onset at each station, marking these times as reference points for subsequent analysis (Figure 14). The detection times from individual stations are evaluated to ensure that an event is not a false alarm, but rather a real earthquake. This is done by applying an evaluation by time intervals. This means that if multiple stations detect an anomaly within a short time interval, it is considered a common event rather than isolated noise or local disturbances. The minimum number of stations to be considered in the evaluation is defined by the user (for more details, see [21] and Supplementary information).

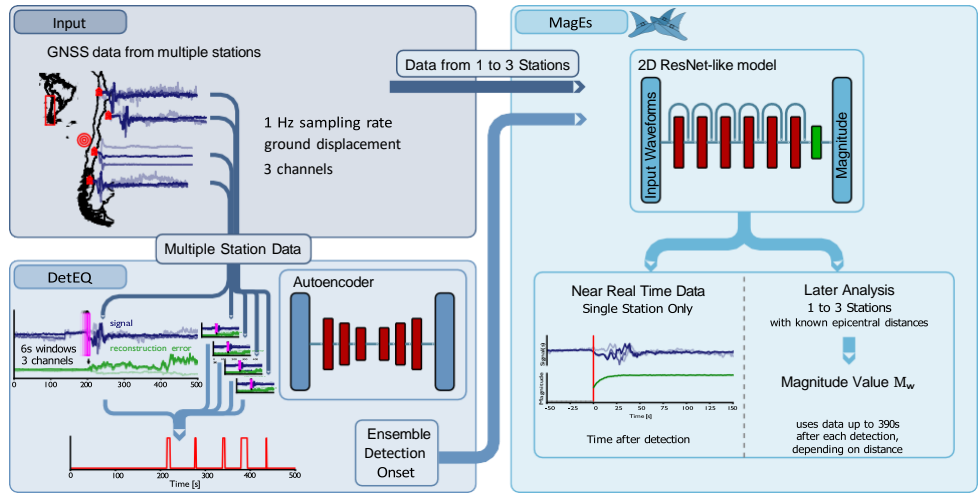

**Fig. 13** Workflow of DetEQ and MagEs models integrated within the SAIPy package. DetEQ performs large earthquake detection using HR-GNSS data from multiple stations, identifying the onset of ground displacements, and considering at least 3 stations. Following detection, MagEs estimate the earthquake magnitude, processing data from one to three stations.

Once DetEQ detects the onset of an event, the corresponding reference times are provided to MagEs for magnitude estimation. In this process, the detected onset time at each station serves as the reference start time for the respective input time series fed into MagEs.

Furthermore, MagEs provides flexibility in data analysis by operating in two modes: a predefined time window mode, used when the epicentral distance is known (see Section 2.2 Input Settings), and a 'time updating' mode, which functions in a near real-time manner. In this mode, MagEs incrementally extends the analysis window second by second, from initial detection to a maximum of 390 seconds post-detection. This continuous updating enables real-time-like magnitude estimation, allowing for progressively refined results as more data become available.

The process of confirming the magnitude output in time updates involves evaluating the consistency of the estimates. Specifically, the model assesses the magnitude for each second and outputs a likely result when the same magnitude value is observed for a defined consecutive period by the user; for example, five seconds (see Figure 15 and Supplementary Information). This means that if the magnitude remains unchanged, for example, for five continuous seconds, it is considered stable and is reported as updated output.

This defined duration can be adjusted according to the specific application, allowing longer or shorter confirmation periods as needed. For fast estimation, the ideal updating period should typically not exceed 60 seconds to ensure a timely response. Also, for reliable output within a short, less than 60-second time window, only the nearest stations are useful (typically within 3 degrees of the epicenter, see Figure 16).

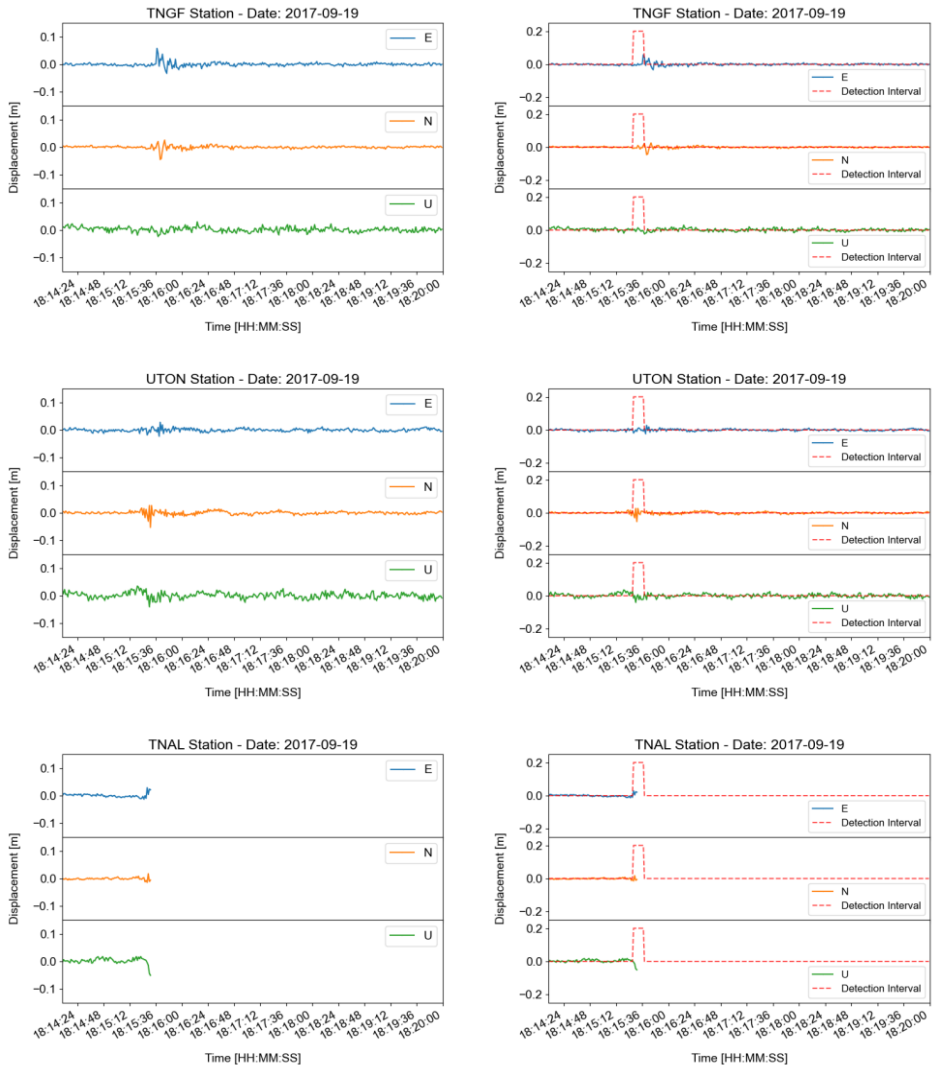

**Fig. 14** Detection by network, using three stations as network threshold criteria, which defined the minimum number of station to confirm the presence of an earthquake. On the left side, the displacement observations from three GNSS stations during the Mw 7.1 Puebla Earthquake on September 19, 2017, are shown. Notably, at the TNAL station, only the initial signal of the earthquake was recorded. On the right side, the detection interval is highlighted, representing the time window in which the onset times from individual stations coincide. These detections meets the predefined network threshold criteria.

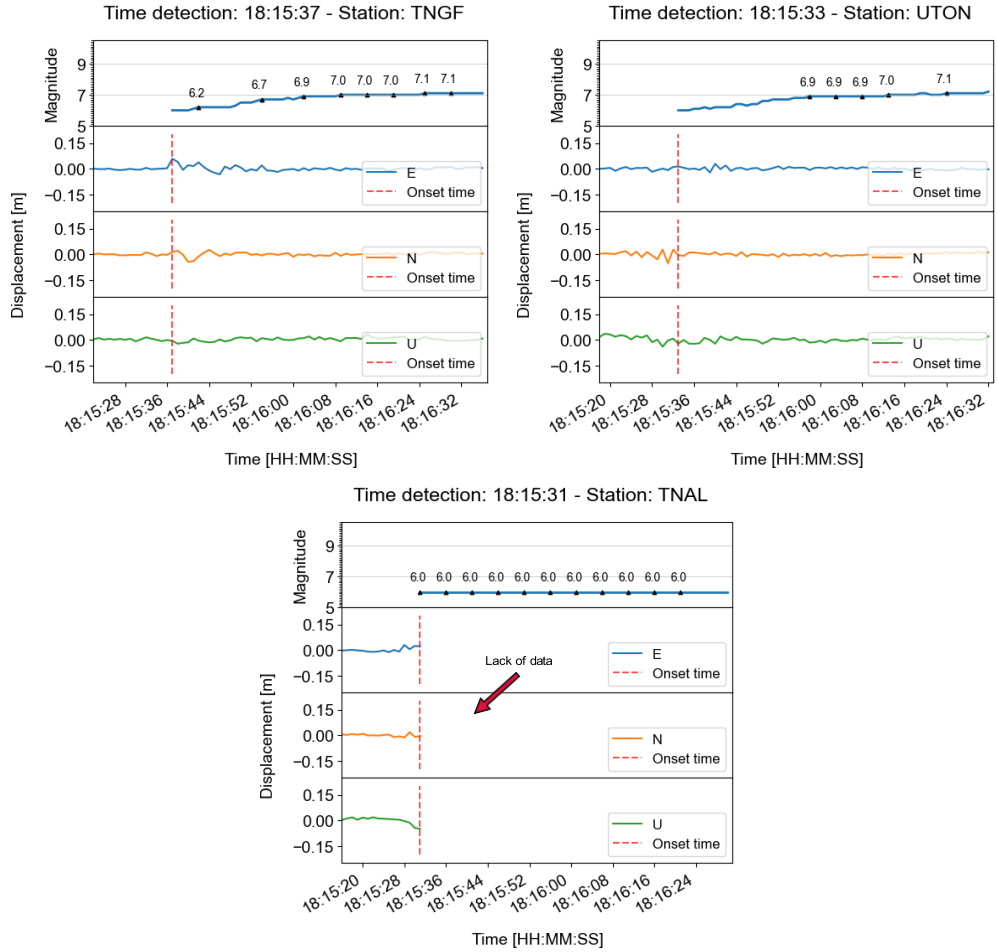

**Fig. 15** Magnitude estimation over 60 seconds after detection using the time updating approach for the three analyzed stations: TNGF, UTON, and TNAL, during the Mw 7.1 Puebla Earthquake. The outputs correspond to magnitudes evaluated with consecutive consistency, with intervals of 5 seconds. The TNGF and UTON stations reach values of Magnitude of 7.1 and 7.0. On the other hand, the TNAL station, which lacks data immediately after detection (Figure 14), results in the minimum magnitude trained for MagEs (Magnitude 6), representing the absence of signal in the time series.

## 4 Conclusions

We have proposed a deep learning model, MagEs, for the magnitude estimation of large earthquakes using HR-GNSS displacement data. The MagEs model is built on an architecture inspired by ResNet, custom made with modifications specific to this particular data analysis. This model effectively captures complex patterns in multi-station time series, enabling it to handle data with variable time window length and number of stations.

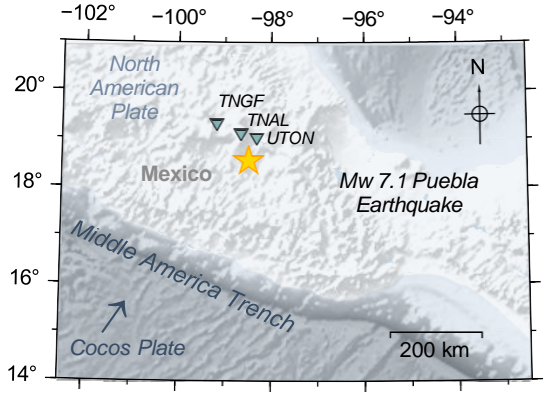

**Fig. 16** Location of the GNSS stations that detected the Mw 7.1 Puebla earthquake using DetEQ. The star symbol indicates the epicenter of the event and the triangles indicate the stations TNGF, UTON, and TNAL, from TLALOCNet.

The model showed robust performance. Under ideal conditions, as demonstrated in tests using synthetic data, the performance of MagEs improves as the number of stations in each set analyzed increases. However, with real data, this trend is not as evident. Inclusion of a station with noisy or low-quality data in a set can introduce errors, affecting overall performance. As a result, the best results are often observed when analyzing single-station data, as this approach makes it easier to recognize if there is an outlier, and thus exclude it from the final result.

Despite being trained on synthetic earthquake data from the Chilean subduction zone, the model demonstrated good performance in most cases, even when tested in other regions (see also the example in the supplementary information). Although earthquakes from diverse tectonic settings generally yielded reliable results, applying transfer learning with data from other regions could further enhance the model performance, particularly for complex cases.

The integration of the proposed model, MagEs, with the DetEQ model [21], within the SAIPy package [12] offers a comprehensive pipeline for seismic analysis with HR-GNSS data. This setup enables continuous monitoring by first detecting an event and then progressively estimating its magnitude as data streams in from multiple stations.

Although the model has not yet been tested in an operational real-time monitoring system, our simulations, where data was streamed and updated second by second, suggest that this approach could be enhanced and adapted in the future for real-time applications.

**Acknowledgements.** This research is supported by the "KI-Nachwuchswissenschaftlerinnen" - grant SAI 01IS20059 by the Bundesministerium für Bildung und Forschung - BMBF. Calculations were performed at the Frankfurt Institute for Advanced Studies GPU cluster, funded by BMBF for the project Seismologie und Artifizielle Intelligenz (SAI).

# Declarations

### Conflict of interest:

The authors have no relevant financial or non-financial interests to disclose.

### Code availability:

All codes in this work are available at https://github.com/srivastavaresearchgroup/SAIPy.

### Data availability:

### Figures, images, and artwork:

The graphic material presented in this paper was made using Matplotlib [8], The Generic Mapping Tools GMT [25], and Inkscape [9].

### Author contribution:

**Claudia Quinteros-Cartaya:** Conceptualization, data processing and preparation, methodology, formal analysis and investigation, validation, writing - original draft preparation, writing - review and editing.
**Javier Quintero-Arenas:** Data processing and preparation, methodology, part of analysis, writing - review and editing.
**Andrea Padilla-Lafarga:** Data processing and preparation, writing - review and editing.
**Carlos Moraila:** Data processing and preparation, writing - review and editing.
**Johannes Faber:** Methodology, writing - review and editing.
**Wei Li:** Methodology, writing - review and editing.
**Jonas Köhler:** Methodology, writing - review and editing.
**Nishtha Srivastava:** Conceptualization, Funding acquisition, Methodology, writing - review and editing.